\documentclass[aps,pra,floatfix,twocolumn,superscriptaddress,floatfix,longbibliography]{revtex4-2}
\usepackage{graphicx}
\usepackage[english]{babel}
\usepackage{amsmath}
\usepackage{amssymb}
\usepackage{tensor}
\usepackage{hyperref}

\newcommand{\vk}{\mathbf{k}}

\newcommand{\vbr}{\hat{\mathbf{r}}}
\newcommand{\be}{\begin{eqnarray}}
\newcommand{\ee}{\end{eqnarray}}
\newcommand{\p}{\partial}

\def\ep#1{\langle #1 \rangle}

\begin{document}

\title{Vortex structure and spectrum of atomic Fermi superfluid in a spherical bubble trap}

\author{Yan He}
\affiliation{College of Physics, Sichuan University, Chengdu, Sichuan 610064, China}
\email{heyan$_$ctp@scu.edu.cn}

\author{Chih-Chun Chien}
\affiliation{Department of Physics, University of California, Merced, CA 95343, USA.}
\email{cchien5@ucmerced.edu}

\begin{abstract}
The structures of multiply quantized vortices (MQVs) of an equal-population atomic Fermi superfluid in a rotating spherical bubble trap approximated as a thin shell are analyzed by solving the Bogoliubov-de Gennes (BdG) equation throughout the BCS-Bose Einstein condensation (BEC) crossover. Consistent with the Poincare-Hopf theorem, a pair of vortices emerge at the poles of the rotation axis in the presence of azimuthal symmetry, and the compact geometry provides confinement for the MQVs. While the single-vorticity vortex structure is similar to that in a planar geometry, higher-vorticity vortices exhibit interesting phenomena at the vortex center, such as a density peak due to accumulation of a normal Fermi gas and reversed circulation of current due to in-gap states carrying angular momentum, in the BCS regime but not the BEC regime because of the subtle relations between the order parameter and density. The energy spectrum shows the number of the in-gap state branches corresponds to the vorticity of a vortex, and an explanation based on a topological correspondence is provided. 
\end{abstract}
\maketitle

\section{Introduction}
Multiply quantized vortices (MQVs), also known as giant vortices, have vorticity higher than the elemental quantum of angular momentum inside. There have been early theoretical predictions~\cite{PhysRevLett.81.2783,PhysRevB.55.11793} and experimental results consistent with MQVs in nano-crystals~\cite{PhysRevLett.107.097202}, mesoscopic~\cite{PhysRevLett.93.257002,KANDA2006122}, and thin-film~\cite{PhysRevLett.103.067007}  superconductors. However, the infinite 2D plane tends to disfavor MQVs because the excitation energy is quadratic in the vorticity~\cite{FetterBook,Pethick_book}, so MQVs tend to decay into multiple single-vorticity  vortices. Nevertheless, confinement effects or symmetries may protect MQVs as observable meta-stable states. There have been more theoretical analyses of MQVs in superconductors~\cite{PhysRevLett.107.057002,Tanaka02,PhysRevLett.119.067003,Vinokur02} and other settings~\cite{Penin2021,PhysRevD.107.026006,PhysRevB.73.235324}.

While quantum vortices have also been studied in cold-atoms (see Refs.~\cite{,Pethick_book,Ueda-book,Tempere17} for reviews), it is challenging to realize MQVs in cold atoms due to the limitation from the conventional harmonic potential. There have been proposals of imposing quadratic plus quartic potentials or other types of potentials to tightly confine the atoms ~\cite{PhysRevA.65.043604,PhysRevA.66.053606,PhysRevLett.92.060401,PhysRevA.74.063619} or utilizing multi-component atomic gases~\cite{PhysRevA.91.043605,PhysRevA.107.053317} to realize MQVs. There have been experimental demonstrations of metastable MQVs in cold bosonic atoms~\cite{PhysRevLett.93.160406,Kumakura06,Okano07,PhysRevA.106.033319}. A different route of realizing MQVs in cold-atoms~\cite{Tomishiyo23,White23} via the recently realized bubble trap~\cite{Carollo22,Lundblad22,TononiReview23,Tononi23} has been proposed for bosonic superfluids, in addition to other theoretical studies of quantum vortices of bosonic superfluids in a spherical geometry~\cite{PhysRevA.102.043305,PhysRevA.103.053306}. Here we explore the structures of MQVs in a fermionic superfluid confined in a spherical bubble trap approximated as a thin shell throughout the BCS-Bose Einstein condensation (BEC) crossover and unravel interesting effects due to the enlarged vortex core and topological properties in the energy spectrum. We mention that  Ref.~\cite{RevModPhys.82.1301} has summarized some properties of vortices on curved surfaces, and here we investigate the structures from a microscopic framework.

By solving the Bogoliubov-de Gennes (BdG) equation~\cite{PhysRevLett.90.210402,PhysRevA.73.041603,PhysRevLett.96.090403} of two-component attractive Fermi gases with equal population in a thin spherical-shell geometry, we characterize the structures of vortices when the gas is rotating about a fixed axis. Our major findings are as follows: (1) Consistent with the Poincare-Hopf theorem~\cite{PHT}, a pair of vortices, one in the north pole and the other in the south pole of the rotation axis, emerge as the rotation exceeds a critical angular velocity. 
(2) For a pair of vortices with vorticity $\nu=\pm 1$ on a sphere, the profiles of the order parameter, density, and superfluid current resemble those of the single vortex in a planar geometry studied previously~\cite{PhysRevLett.90.210402,PhysRevA.73.041603,PhysRevLett.96.090403}. (3) For a pair of vortices with $\nu=\pm 2$, a normal Fermi gas may over-occupy the enlarged vortex core in the BCS regime, leading to a density peak, not a density dip, at the center of the vortex core. In contrast, the regular density dip remains robust in the BEC regime for the $\nu=2$ vortex. 
(4) For a pair of $\nu=\pm 3$ vortices on a sphere in the BCS regime, a reversed circulation of the super-current emerges inside the vortex core, in addition to the density peak due to the occupation of a normal Fermi gas in the core. (5) For a MQV with vorticity $\nu$, there are $\nu$ branches of in-gap states in the energy spectrum, which are argued to be from a topological origin.

In the literature, consistency with the Poincare-Hopf theorem has been discussed in bosonic superfluids in spherical bubble traps~\cite{PhysRevA.102.043305}, $p$-wave superfluids on a sphere~\cite{Fan16}, and the classical XY model on a spherical lattice~\cite{PhysRevResearch.4.023005}. Possibilities of a reversed super-current inside a MQV have been speculated in Ref.~\cite{PhysRevLett.119.067003}, where a suppression of the angular momentum in BCS MQVs is discussed, and in Ref.~\cite{PhysRevA.106.033322} discussing  vortex structures of population-imbalanced Fermi superfluids. The MQVs of equal-population Fermi superfluids in spherical bubble traps presented here will offer a feasible way for investigating the intriguing phenomenon of counter-circulating current inside a giant vortex. Moreover, we will provide an explanation of the topological correspondence between the vorticity and the number of in-gap states via an analogy with the Chern insulator.

The rest of the paper is organized as follows. Sec.~\ref{sec:theory} presents the BdG equation of an atomic Fermi superfluid in a spherical bubble trap approximated by a thin shell and its vortex solutions. Physical quantities such as the gap function, density, and current for characterizing vortices are introduced. Sec.~\ref{sec:result} presents the vortex solutions with vorticity $\nu=1,2,3$ and explains the density peak and reversed circulation of current at the core center for higher-vorticity vortices in the BCS regime. The energy spectrum showing the in-gap states will be shown and the topological correspondence between the in-gap states and vorticity will be explained. Some implications for experimental studies are also discussed. Sec.~\ref{sec:conclusion} concludes our work.

\section{Theoretical background}\label{sec:theory}

\subsection{BdG equation on sphere}
To investigate the vortices of a fermionic superfluid in a bubble trap throughout the BCS-BEC crossover at $T=0$, we setup and solve the BdG equation on a sphere. Here we set $\hbar=1$ and $k_B=1$. 
On a thin spherical shell of radius $R_0$, the single-particle Hamiltonian in the spherical coordinates can be expressed as
\be
H_0=-\frac{1}{2M}\Big[\frac{1}{r^2}\frac{\p}{\p r}r^2\frac{\p}{\p r}+\frac{1}{r^2}\nabla_s^2\Big]+V(r-R_0).
\ee  
Here $M$ is the fermion mass and $\nabla_s^2$ is the spherical Laplacian to be explained later. The spherical-shell potential $V(r-R_0)$ is assumed to be highly concentrated at $r=R_0$, thus the $r$ derivative terms can be ignored. The single fermion Hamiltonian then becomes
\be
H_0=-\frac{1}{2MR_0^2}\nabla_s^2.
\ee
In the following, we take $\mathcal{E}_0=\frac{1}{2MR_0^2}$ as the energy unit and will no longer show it explicitly. The spherical Laplacian operator is given by 
\be
\nabla_s^2&\equiv&-\frac{1}{\sqrt{G}}\p_\mu\sqrt{G}G^{\mu\nu}\p_\nu \nonumber \\
&=&-\Big(\frac{1}{\sin\theta}\frac{\p}{\p\theta}\sin\theta\frac{\p}{\p\theta}+\frac{1}{\sin^2\theta}\frac{\p^2}{\p^2\phi}\Big).
\ee
where $G_{\mu\nu}=\mathrm{diag}\{1,\sin^2\theta\}$ is the metric on a 2D unit sphere and $G=\det(G_{\mu\nu})$. By keeping the radius $R_0$ implicit, the fermion system is effectively on a unit sphere characterized by the spherical coordinates $\vbr=(\theta,\phi)$.

Following Ref.~\cite{He22}, a fermionic superfluid on a thin spherical shell is described by the BCS mean-field Hamiltonian given by
\be
H_{\mathrm{BCS}}&=&\int_{S^2}d\vbr\Big[\sum_{\sigma}\psi_{\sigma}^{\dag}(\vbr)\hat{T}\psi_{\sigma}(\vbr)
+\Delta(\vbr)\psi^{\dag}_{\uparrow}(\vbr)\psi^{\dag}_{\downarrow}(\vbr)+ \nonumber \\
& &\Delta^*(\vbr)\psi_{\downarrow}(\vbr)\psi_{\uparrow}(\vbr)\Big].
\label{eq-BCS}
\ee
Here the surface element is $d\vbr=\sin\theta d\theta\phi$. $\psi_{\sigma}$ is the fermion annihilation operator of spin $\sigma=\uparrow,\downarrow$. The kinematic energy operator is given by
\be
\hat{T}=H_0-\mu.
\ee
Here we assume equal population of the two components. The gap function representing the order parameter is defined as
\be
\Delta(\vbr)=-g\ep{\psi_{\uparrow}(\vbr)\psi_{\downarrow}(\vbr)},
\ee
where $g$ is the bare coupling constant.

The BCS mean-field Hamiltonian can be diagonalized by the Bogoliubov transformation, which introduces the quasi-particle operators as
\be
\psi_{\uparrow}(\vbr)=\sum_n\Big[u_n(\vbr)\gamma_{n,1}-v^*_n(\vbr)\gamma^\dag_{n,2}\Big],\nonumber\\
\psi^{\dag}_{\downarrow}(\vbr)=\sum_n\Big[v_n(\vbr)\gamma_{n,1}+u^*_n(\vbr)\gamma^\dag_{n,2}\Big].
\label{eq-Bogo}
\ee
The coefficients of the above transformation satisfy the orthonormal conditions
\be
\int_{S^2}d\vbr\Big[u^*_m(\vbr)u_n(\vbr)+v^*_m(\vbr)v_n(\vbr)\Big]=\delta_{mn}.
\ee
In terms of $\gamma$, the BCS Hamiltonian can be expressed as
\be
H_{\mathrm{BCS}}=E_0+\sum_{n,\sigma}E_n\gamma^{\dag}_{n,\sigma}\gamma_{n,\sigma},
\ee
where $E_0$ is the ground-state energy. With the above Hamiltonian, one has the following commutation relations 
\be
[H_{\mathrm{BCS}},\gamma_{n,\sigma}]=-E_n\gamma_{n,\sigma},\quad
[H_{\mathrm{BCS}},\gamma^{\dag}_{n,\sigma}]=E_n\gamma^{\dag}_{n,\sigma}.
\ee
Substituting Eq.~(\ref{eq-BCS}) and the inverse of Eq.~({\ref{eq-Bogo}}) into the above equations and equating both sides, we arrived at the Bogoliubov-de Gennes equation of a Fermi superfluid in a spherical-shell geometry. Explicitly, 
\be
\left(
  \begin{array}{cc}
    \hat{T} & \Delta(\vbr) \\
    \Delta^*(\vbr) & -\hat{T}
  \end{array}
\right)\left(
  \begin{array}{c}
    u_n(\vbr) \\
    v_n(\vbr)
  \end{array}
\right)=E_n\left(
  \begin{array}{c}
    u_n(\vbr) \\
    v_n(\vbr)
  \end{array}
\right).
\ee
The gap function is determined by the wave functions $u_n$ and $v_n$ as
\be
\Delta(\vbr)=g\sum_n u_n(\vbr)v^*_n(\vbr)\Big[1-2n_F(E_n)\Big].
\label{eq-del}
\ee
Here $n_F(x)=1/(e^{x/T}+1)$ is the Fermi distribution function. The summation $\sum_n$ is for eigen-energies that satisfy the condition $0\le E_n\le E_{\textrm{cut}}$ with some cutoff energy $E_{\textrm{cut}}$. Here we assume that the cutoff energy is the largest kinematic energy $E_{\textrm{cut}}=L_{max}(L_{max}+1)$ with the maximal angular momentum quantum number $L_{max}$. Further increments of $E_{cut}$ do not lead to qualitative changes of the results. More details are given below in the discussion of vortex solutions. 

The bare coupling constant $g$ is related to the two-body scattering length $a$ by making use of the following renormalization relation on a thin spherical shell~\cite{He22}:
\be\label{eq:ganda}
&&\frac 1g=\int dl\frac{2l+1}{2\epsilon_l+|E_b|}=\int_0^{E_{\textrm{cut}}}\frac{d\epsilon_l}{2\epsilon_l+|E_b|}.
\ee
Here $\epsilon_l=l(l+1)$ is the dispersion of free fermions on a unit sphere and $E_b=-\frac{1}{Ma^2}$ is the binding energy of the two-body bound state. Since the two-body scattering length in 2D is positive, solving Eq.~\eqref{eq:ganda} allows us to translate the coupling constant $g$ to the dimensionless parameter $-\ln(k_Fa)$, which varies from negative to positive values throughout the BCS-BEC crossover. The BCS-BEC crossover occurs when the chemical potential changes sign~\cite{He22}, implying a change of nature of the Fermi superfluid as the attraction increases.
Following Ref.~\cite{He22},  there are two ways to tune the ratio between the kinetic and interaction energies of a Fermi superfluid on a thin spherical shell and push the system through the BCS-BEC crossover. One may tune the scattering length $a$ to increase the pairing strength or increase the size of the spherical shell to lower the Fermi energy. As the size of the sphere increases, the curvature decreases, leading to a curvature-induced crossover of atomic Fermi superfluids. A combination of the two approaches will make it more feasible to study the whole crossover of a Fermi superfluid in a spherical bubble trap approximated as a thin shell.

\subsection{Vortex solutions}
\begin{figure}
\centering
\includegraphics[width=0.4\columnwidth]{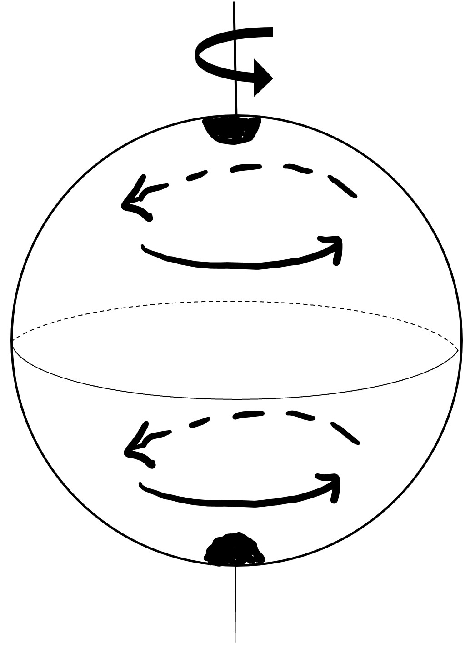}
\caption{\label{fig:illustration} Illustration of a Fermi superfluid on a 2D thin spherical shell under rotation (indicated by the circular arrow around the rotation axis). A pair of vortices (the black dots) emerge at the north and south poles while the currents (black arrows on the sphere) circulate the vortices.
}
\end{figure}

We consider a solution of the BdG equation with a pair of vortices located at the north and south poles of a thin spherical shell, where the rotation axis goes through the two poles. Figure~\ref{fig:illustration} schematically shows the setup. The azimuthal symmetry is assumed to hold. For vortices with vorticity $\nu$, the gap function is assumed to take the following form
\be
\Delta(\vbr)=\Delta(\theta)e^{-i\nu\phi},
\ee
which means the vorticities of the two vortices are $\pm\nu$, respectively. With the functional form of the gap function, we can expand $u_n(\vbr)$ and $v_n(\vbr)$ by the spherical harmonics as
\be
&&u_n(\vbr)=\sum_{l,m}c_{nlm}Y_{l,m}(\theta,\phi),\\
&&v_n(\vbr)=\sum_{l,m}d_{nlm}Y_{l,m+\nu}(\theta,\phi).
\label{eq-uv}
\ee
Then the BdG equation becomes a matrix eigenvalue equation. However, the matrix can be split into diagonal blocks for different values of $m$. For a given $m$, the BdG equation can be written as
\be
&&\sum_{ll'}\left(
  \begin{array}{cc}
    T_m & D_m \\
    D_m^T & -T_{m+1}
  \end{array}
\right)_{ll'}\left(
  \begin{array}{c}
    c_{nl'm} \\
    d_{nl'm}
  \end{array}
\right)=E_n\left(
  \begin{array}{c}
    c_{nlm} \\
    d_{nlm}
  \end{array}
\right).\label{eq-BdG}
\ee
Since the spherical harmonics are the eigen-functions of $\nabla^2_s$, we find that the kinematic term is simply a diagonal matrix given by
\be
(T_m)_{ll'}=\Big(\frac{l(l+1)}{2M}-\mu\Big)\delta_{ll'},~l=m,m+1,\cdots,L_{max.}
\ee
The matrix elements of the gap function require integrations of the form 
\be
(D_m)_{ll'}=\int_{-1}^1dx\Delta(x)N_{l,m}P^m_l(x)N_{l',m+1}P^{m+1}_{l'}(x).
\ee
Here we have used the definition of spherical Harmonics to introduce the associated Legendre polynomial $P^m_l(x)$ with $x=\cos\theta$, the normalization factor is $N_{l,m}=\sqrt{\frac{(l-|m|)!}{(l+|m|)!}\frac{2l+1}{2}}$, and the $\phi$ integral has been completed.

In our numerical computations, the integration is calculated by Gaussian quadrature. For the present case, it has the form 
\be
\int_{-1}^1 h(x)dx=\sum_i w_i h(z_i),
\ee
where $z_i$ are the zeros of the Legendre polynomial $P_{L_M}(x)$ with large enough $L_{M}$. Here the weights are given by  $w_i=\dfrac{2}{(1-x_i^2)[P'_{L_M}(x_i)]^2}$. 
Since the integration functions of this work are polynomials of order smaller than $2L_{max}$, the above Gaussian quadrature will be accurate as long as we take $L_{M}>2L_{max}$.
Using Eq.~(\ref{eq-BdG}), the BdG equation has been transformed into a set of matrix eigen-equations of dimension $2(L_{max}-|m|+1)$  for each $m$. We can diagonalize each matrix corresponding to a specific $m$ to find the eigen-energies $\{E_n\}$ and the eigen-functions $\{u_n(x)\}$ and $\{v_n(x)\}$. This procedure will be repeated for all possible $m$ from $-L_{max}$ to $L_{max}$. We take $L_{max}=30$ in our numerical calculations since further increments of $L_{max}$ do not lead to observable changes in the results. In the numerical calculations, the fermion mass is set to $M=1/2$ for simplicity. 

To obtain a self-consistent solution of the BdG equation, we start with an initial guess of the gap function, which vanishes at both the north and south poles of the sphere. For example, one may try 
$\Delta(x)=\Delta_0(1-x^2)$ as an initial guess. This is consistent with the boundary conditions of the two vortices located at the poles of the sphere.
From Eq.~(\ref{eq-BdG}), we find the eigen-functions $u_n(x)$ and $v_n(x)$ and generate a new gap function $\Delta'(x)$ according to Eq.~(\ref{eq-del}) to be used for the next round. These steps will be repeated many times until the convergence of $\Delta(x)$ is reached. The convergence condition that we used is 
\be
\int_{-1}^1\Big|\Delta'(x)-\Delta(x)\Big|dx\Big/\int_{-1}^1|\Delta(x)|dx<\epsilon,
\ee
where we have taken  $\epsilon=10^{-3}$.
During the iteration for obtaining a self-consistent solution, the chemical potential $\mu$ is kept fixed. 

In order to make a comparison with the uniform case, we determine the Fermi energy $E_F$, which is the Fermi energy of a free Fermi gas with the same total particle number in the same geometry. The fermion density of both spins is given by
\be
n(\vbr)=2\sum_n \Big[|u_n(\vbr)|^2 n_F(E_n)+|v_n(\vbr)|^2(1-n_F(E_n))\Big].\nonumber\\
\label{eq-num}
\ee
Integrating Eq.~(\ref{eq-num}) over the sphere gives the total particle number $N$ of the Fermi superfluid. Assuming that the same amount of free fermions on the same sphere fills up to angular quantum number $L$, then we have
\be
N=2\int_0^L (2l+1)dl=2L(L+1).
\ee
Therefore, the Fermi energy on a sphere is given by $E_F=L(L+1)=N/2$, and the corresponding Fermi momentum is $k_F=\sqrt{2ME_F}$. We will normalize the results by $E_F$ and $k_F$.

Since a vortex is accompanied by a circulating current, we also evaluate the particle current, which can be obtained by 
\be
\mathbf{J}(\vbr)=\frac{i}{2M}\sum_{\sigma}\ep{\psi^{\dag}_{\sigma}(\vbr)\nabla\psi_{\sigma}(\vbr)-\nabla\psi^{\dag}_{\sigma}(\vbr)\psi_{\sigma}(\vbr)}.
\ee
Making use of the Bogoliubov transformation of Eq.~(\ref{eq-Bogo}) and the $\phi$ dependence in Eq.~(\ref{eq-uv}), we find that the circulating current can be computed from $u_n$ and $v_n$ as
\be
\mathbf{J}(\vbr)&=&\frac{2}{M\sin\theta}\sum_{n,m} \Big[m|u_n(\vbr)|^2 n_F(E_n)- \nonumber \\
& &(m+1)|v_n(\vbr)|^2(1-n_F(E_n))\Big]\mathbf{e}_{\phi}.
\ee
We note that the BdG equation allows spatial resolution of all the physical quantities, including the density and current, which will reveal interesting physics inside the MQVs.
The numerical results of the MQV will be presented in the next section.

The energy of a vortex $E_v$ is usually higher than the ground-state energy $E_0$ of a uniform state~\cite{FetterBook,Pethick_book} because the vortex states are  excitations of the underlying superfluid. Moreover, he excitation energy of a vortex with vorticity $\nu$ is proportional to $\nu^2$~\cite{FetterBook,Pethick_book}, therefore a MQV tends to decay into multiple vortices with lower vorticity. To stabilize the vortex state, however, one may rotate the whole system~\cite{Tomishiyo23}, which is equivalent to adding an extra term to the kinematic operator given by
\be\label{eq:TwL}
\hat{T}'=H_0-\mu-\Omega \hat{L}_z.
\ee  
Here $\hat{L}_z=i\frac{\p}{\p\phi}$ and $\Omega$ is the angular velocity of the rotation. Therefore, in the laboratory frame, the energy of the vortex state is
$E'_v=E_v-\Omega\ep{\hat{L}_z}$.
Therefore, if the angular velocity exceeds the following bound
$\Omega>\frac{E_v-E_0}{\ep{\hat{L}_z}}$,
then $E'_v<E_0$ in the laboratory frame. Here $\ep{\hat{L}_z}$ is proportional to $\nu$. Therefore, for large enough $\Omega$ and $\nu$, the MQV states may be stabilized, as mentioned in Refs.~\cite{Tomishiyo23,White23}.    Additionally, cold-atoms in spherical bubble traps are promising to maintain the MQVs due to the strong confinement from the compact geometry and the azimuthal symmetry disfavoring multi-vortex configurations violating the symmetry. More discussion will be presented later. We mention that our study considers the sphere to be much larger than the vortices, so the curvature only tunes the BCS-BEC crossover and does not substantially affect the stability of the vortices.

\begin{figure}
\centering
\includegraphics[width=0.75\columnwidth]{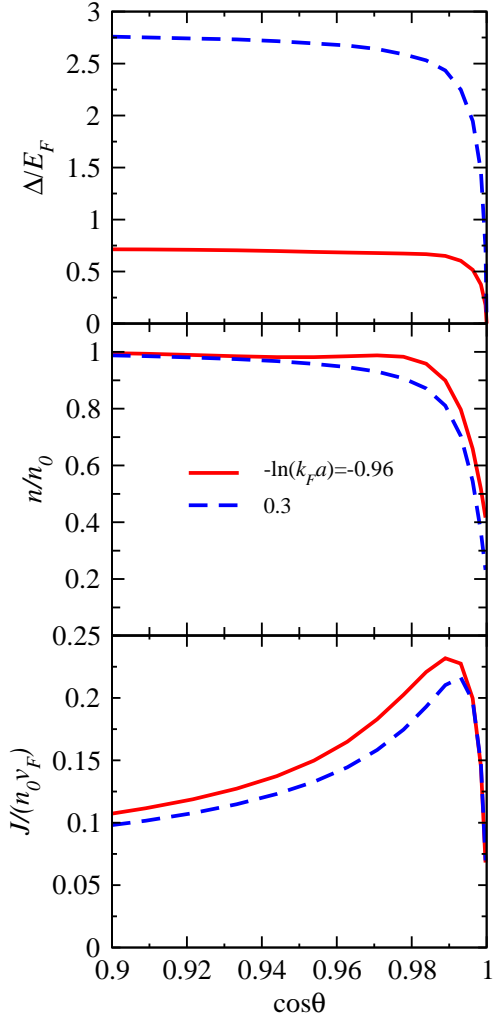}
\caption{The pairing gap $\Delta$, density $n$, and particle current $J$ as a function of $\cos\theta$ for $\nu=1$ vortices in the BCS regime (red full line) and BEC regime (blue dashed line). The BCS and BEC cases correspond to $-\ln(k_F a)=-0.96$ and $0.3$. The corresponding $\mu/E_F=0.9$ and $-0.4$. Here $n_0=N/(4\pi)$ is the averaged density on the sphere while $E_F$ and $v_F$ are the Fermi energy and Fermi velocity of a noninteracting Fermi gas with the same geometry and particle number.}
\label{del-dens}
\end{figure}

\section{Result and discussion}\label{sec:result}

\subsection{Vortex structures}
For a pair of $\nu=\pm1$ vortices on a sphere, the numerical results of the gap function $\Delta(x)$, density $n(x)$, and current $J(x)$ are shown in Figure \ref{del-dens} for two selected cases: $-\ln(k_F a)=-0.96$ with $\mu/E_F=0.9$ and $-\ln(k_F a)=0.3$ with $\mu/E_F=-0.4$. Here $x=\cos\theta$. The positive (negative) value of $\mu$ confirms the Fermi superfluid is in the BCS (BEC) regime. The two cases are chosen because they are not far away from the crossover indicated by $\mu=0$ and exhibit some contrasting properties of BCS and BEC superfluids. We caution that different pairing strengths tune the interaction energy scale, but the stability of the vortices are determined by the angular velocity, confining potential, vortex energy, and vorticity. Therefore, the vortices may be stabilized in different regimes of the BCS-BEC crossover illustrated here by suitable choices of the parameters.

While the order parameter represented by the gap function vanishes in the vortex center, the depletion of the density inside the vortex increases as the system getting deeper into the BEC regime. Finally, the circulating current shows a maximum, which indicates the size of the vortex. The results are similar to the $\nu=1$ vortex on a 2D plane studied previously~\cite{PhysRevLett.90.210402,PhysRevA.73.041603,PhysRevLett.96.090403}.

We comment on a subtlety about the Poincare-Hopf theorem when applied to a rotating superfluid on a 2D surface. The Poincare-Hopf theory concerns the total index of the singularities of a tangent vector field on a surface. Meanwhile, the vortex results from the phase field of the order parameter, which at first look is not a tangent vector field in real space. Nevertheless, the circulating current $\mathbf{J}$ reflects the gradient of the phase field~\cite{FetterBook,Pethick_book} and forms a tangent vector field on the surface. Therefore, the Poincare-Hopf theorem constrains the net number of vortices of a superfluid confined on a surface due to the underlying topology in a subtle way.

\begin{figure}
\centering
\includegraphics[width=0.75\columnwidth]{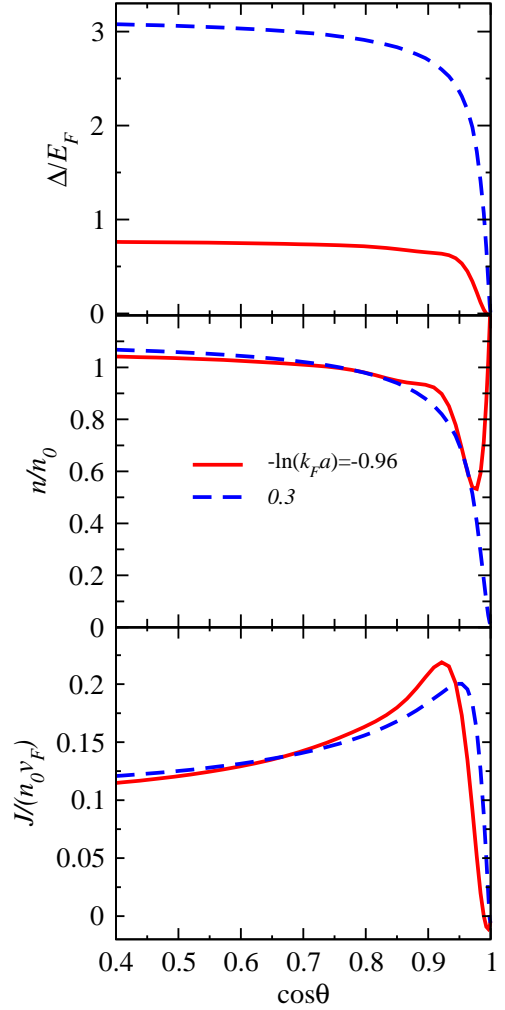}
\caption{The gap function $\Delta$, density $n$, and particle current $J$ as functions of $\cos\theta$ for a $\nu=2$ vortex. The red solid and blue dashed lines correspond to $-\ln(k_F a)=-0.96$ and $0.3$ with $\mu/E_F=0.9$ and $-0.4$, respectively.}
\label{vor-2}
\end{figure}

\begin{figure}
\centering
\includegraphics[width=0.75\columnwidth]{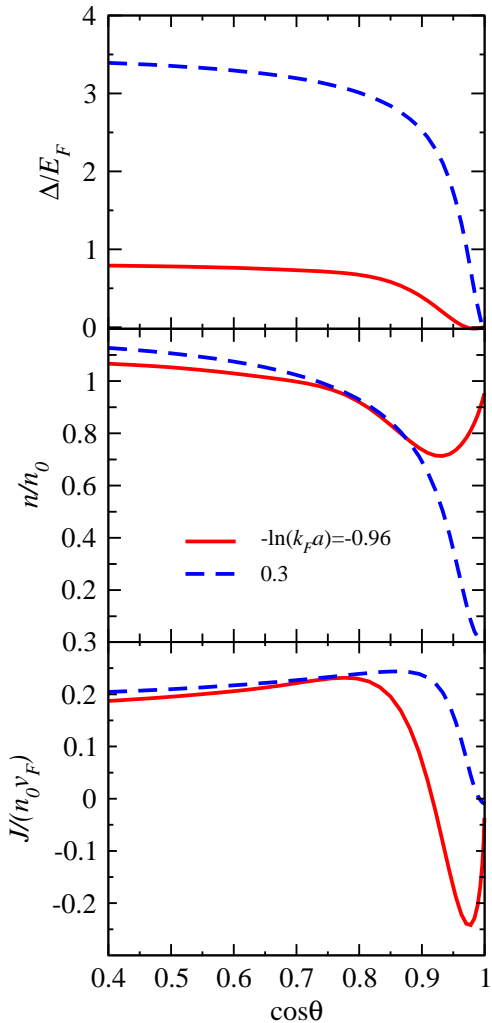}
\caption{Profiles of the gap function $\Delta$, density $n$, and particle current $J$ as functions of $\cos\theta$ for a $\nu=3$ vortex. The solid (dashed) line corresponds to $-\ln(k_F a)=-0.96$ ($-\ln(k_F a)=0.3$) with $\mu/E_f=0.9$ ($\mu/E_f=-0.4$) in the BCS (BEC) regime.
}
\label{vor-3}
\end{figure}

For a pair of $\nu=\pm 2$ vortices on a sphere, the numerical results of the gap function $\Delta(x)$, density $n(x)$, and current $J(x)$ are shown in Figure \ref{vor-2} for the same selected cases $-\ln(k_F a)=-0.96$, and $0.3$. The chemical potential is negative (positive) for the first (second) case, indicating the Fermi superfluid is in the BCS (BEC) regime. The results in the BEC regime look similar to those of $\nu=1$ in the BEC regime except for the larger size of the vortex and the more complete depletion of the density at the vortex center. In contrast, the density of a $\nu=2$ vortex shows a density peak instead of a dip at the center of the vortex in the BCS regime when $\mu$ is close to $E_F$.

The density peak inside a MQV in the BCS regime can be understood by a qualitative argument from the BdG equation. Since the chemical potential is positive in the BCS regime while the gap function vanishes inside the vortex core, a normal Fermi gas survives inside the vortex core. For the $\nu=1$ case, this leads to incomplete depletion of the density inside the core. For MQVs with higher $\nu$, the enlargement of the vortex core allows more normal fermions in the BCS regime to accumulate there and eventually give rise to a density peak at the core center. The accumulation of the normal Fermi gas is not possible in the BEC regime because the strong pairing of fermions forms composite bosons, which fix the order parameter with the density and lead to a negative chemical potential of the fermions. Therefore, the density peaks of MQVs in the BCS regime contribute to features not observable for bosonic superfluids in the same setup.

Figure \ref{vor-3} shows the results of a vortex with $\nu=3$ for the two selected cases in the BCS and BEC regimes. Due to its high vorticity, the vortex core is even larger. The results in the BEC regime still resemble the BEC results of the $\nu=1, 2$ cases. In the BCS regime, the density again shows a peak at the vortex center due to the accumulation of normal fermions. However, another feature emerges in the core, where one can see that the circulation of the current is reversed near the vortex center as indicated by the negative value of the current. The result thus confirms the conjecture of Ref.~\cite{PhysRevLett.119.067003} that reversed circulation may reside in the cores of higher MQVs of Fermi superfluids. In contrast to Ref.~\cite{PhysRevA.106.033322}, where reversed circulation was discussed in the vortex core of  population-imbalanced Fermi superfluids, here we show that for an equal-population Fermi superfluid in a spherical bubble trap, reversed circulation may be prominent in MQVs with $\nu\ge 3$. A careful examination of the current in the core of the $\nu=2$ vortex shown in Fig.~\ref{vor-2} reveals that a slight reversal of the current already occurs at the center in the BCS regime. However, the tiny region and magnitude of the reversed current of a $\nu=2$ vortex suggests it is more feasible to investigate vortices with $\nu\ge 3$ to probe the phenomenon. As we will show by analyzing the energy spectrum from the BdG equation in the next subsection, the reversed current is associated with the states in the core that carry angular momentum.

\begin{figure}
\centering
\includegraphics[width=\columnwidth]{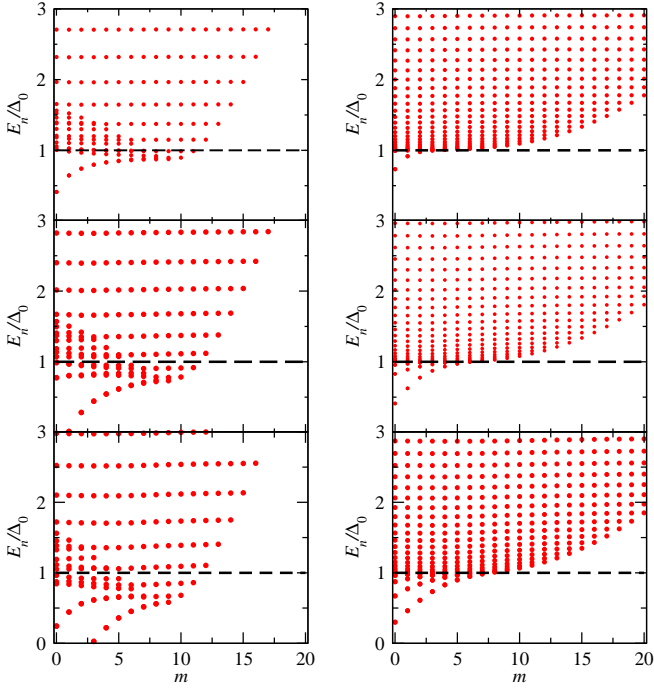}
\caption{From top to bottom: The eigenvalues $E_n$ as a function of $m$ from the BdG equation of vortices with vorticity $\nu=1,2,3$. The left (right) column is the BCS (BEC) case with $-\ln(k_F a)=-0.96$  ($-\ln(k_F a)=0.3$). The black dashed lines represent the value of the bulk gap $\Delta_0$. For the left (right) column, $\Delta_0/E_F=0.7$ ($\Delta_0/E_F=2.8$).
}
\label{eigen}
\end{figure}

\subsection{Energy spectrum and in-gap states}
The eigen-energy spectrum of the vortex solutions from the BdG equation is shown in  Fig.~\ref{eigen}. There are in-gap states with $E_n < \Delta_0$ in every case, where $\Delta_0$ is the gap function away from the vortex. We found that the number of in-gap states increases with the vorticity. What is interesting here is that for higher-vorticity
vortices ($\nu=2, 3$ for example), the in-gap states near $E=0$ start acquiring finite angular momentum in the BCS regime, as indicated by the $m>0$ eigenvalues near the zero energy. Those states near $E=0$ will be shown to be inside the vortex core and carry angular momentum. They  lead to the counter-circulation of the current inside the vortex core of a vortex with higher vorticity, as shown in Fig.~\ref{vor-3} and conjectured in Ref.~\cite{PhysRevLett.119.067003}.

Figure \ref{eigen} suggests that the number of the branches of the in-gap states is equal to the vorticity of the MQV. Explicitly, for a vortex with vorticity $\nu$, there are $\nu$ branches of in-gap states in its spectrum.
Here we provide a heuristic argument to support this correspondence. We note that the BdG equation in momentum space may be viewed as a two-band model. In the first-quantization language, it can be expressed as
\be
H=\textrm{Re}(\Delta)\,\sigma_1+\textrm{Im}(\Delta)\,\sigma_2+(\frac{\vk^2}{2M}-\mu)\sigma_3.
\label{eq-band}
\ee
Here $\sigma_{1,2,3}$ are the Pauli matrices.
For a MQV with vorticity $\nu$, we assume $\Delta=f(\theta)e^{i\nu\phi}$, where the amplitude $f(\theta)$ satisfies the boundary condition $f(0)=f(\pi)=0$. The Chern number of the above 2D two-band model can be obtained as follows. Since quantized topological indices do not depend on the details of the functional forms, we may assume that $f(\theta)=\sin^{\nu}\theta$ for simplicity. By defining $k_{\pm}=k_x\pm ik_y=\sin\theta e^{\pm i\phi}$, the two-band mode of Eq.~(\ref{eq-band}) then becomes
\be
H=\textrm{Re}k_{+}^{\nu}\,\sigma_1+\textrm{Im}k_{+}^{\nu}\,\sigma_2+ \left(\frac{\vk^2}{2M}-\mu\right)\sigma_3.
\label{eq-band1}
\ee
For $\nu=1$, the above model is the continuum limit of the Qi-Wu-Zhang model of Chern insulator \cite{Qi2} with the Hamiltonian 
\be
H_{QWZ}&=&\sin k_x\sigma_1+\sin k_y\sigma_2+ \nonumber \\
& &\Big(\frac{2-\cos k_x-\cos k_y}M-\mu\Big)\sigma_3.
\ee
It is known that, for $\mu$ not too large, the Qi-Wu-Zhang model has Chern number $C=1$, which is the same as the vorticity $\nu=1$ of the vortex solution from the BdG equation. Thus, a connection between the vortex solution of the BdG equation and the Chern insulator has been built.

For higher values of $\nu$, we construct a 3D vector with the components at small $\mathbf{k}$ corresponding to the coefficients of the Hamiltonian. Explicitly,
\be
\textbf{R}=\Big(\textrm{Re}k_{+}^{\nu},\textrm{Im}k_{+}^{\nu},\frac{2-\cos k_x-\cos k_y}M-\mu\Big).
\ee
The Chern number is then given by
\be
C=\frac{1}{4\pi}\int d^2k\frac{\textbf{R}\cdot\p_x\textbf{R}\times\p_y\textbf{R}}{R^3}
\ee
with $R=|\textbf{R}|$ and $\p_i=\frac{\p}{\p k_i}$. For $\mu$ not too large, one can verify that $C=\nu$, which extends the connection between the vortex solutions from the BdG equation and Chern insulator. According to the bulk-edge correspondence of the Chern insulator~\cite{Kane_TIRev,Zhang_TIRev,Chiu2016}, the number of edge modes located in the gap between the two bands should be the same as the Chern number. In the vortex solutions on a sphere, the cores of the two vortices at the north and south poles support localized states analogous to the edge modes inside the band gap. Therefore, the number of the in-gap state branches is equal to the vorticity that plays the role of the Chern number.

We mentioned that there have been discussions of Dirac fermions inside a scalar vortex~\cite{JACKIW1981681,PhysRevB.107.125418,Gates23}, where it was suggested that for a MQV with vorticity $\nu$, the Dirac fermions will support $\nu$ zero modes. By viewing the pairing gap $\Delta$ as a dynamically generated scalar field, our results provide another example connecting the topological indices of MQVs with interesting states of fermions inside their cores.

\begin{figure}
\centering
\includegraphics[width=\columnwidth]{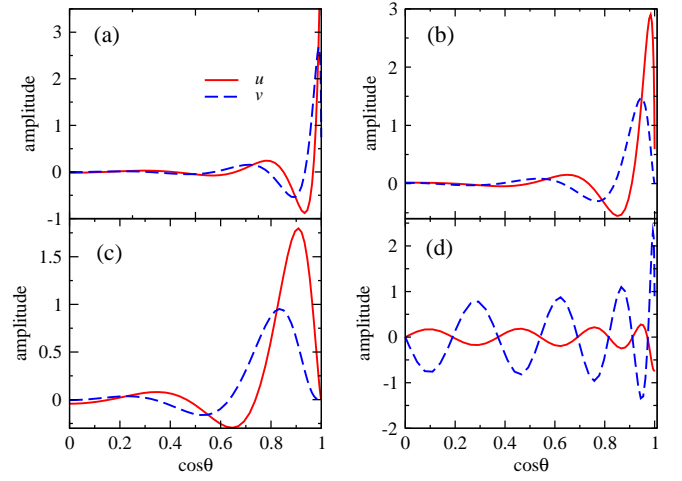}
\caption{The BdG eigen-functions $u$ (solid lines) and $v$ (dashed lines) as functions of $\cos\theta$ of the vortex solutions in the BCS regime with $-\ln(k_F a)=-0.96$. Panels (a), (b), and (c) show the lowest-energy in-gap states for the solutions with vorticity $\nu=1,\,2,\,3$,  respectively. Panel (d) shows a typical bulk state of $\nu=1$.
}
\label{uv}
\end{figure}

Fig.~\ref{uv} shows the eigenfunctions $u_n$ and $v_n$ of selected in-gap states with lowest energies and a typical bulk state with $E > \Delta_0$ in the BCS case. One can see that the in-gap states are indeed localized inside the vortex core while the bulk states extends to the whole sphere. Here the lowest energy in-gap states for the vortices with  vorticity $\nu=1,2,3$ have $m=0,1,3$, respectively. As $\nu$ increases, the peaks of $u$ and $v$ are moving away from the center because of the enlarged vortex core. 
For higher-vorticity vortices in the BCS regime, those in-gap states with finite angular momentum  contribute to the reversed circulation at the core center. 
The in-gap states in the BEC regime are more localized compared to those in the BCS regime and do not carry significant angular momentum. Nevertheless, the eigenfunctions in the BEC regime are qualitatively similar to those in the BCS regime shown in Fig.~\ref{uv}.

\subsection{Implications}
The reason that interesting behavior, such as the density peak or reversed circulation of current at the vortex center in the BCS regime but not BEC regime, is due to the different relations between the density and gap function in the BCS-BEC crossover. In the BCS regime, the density and gap function are two different quantities, and a vanishing gap does not imply zero density. Therefore, a normal Fermi gas is allowed to enter the vortex core with a vanishing order parameter due to the positive chemical potential and gives rise to a density peak at the core center for higher MQVs. In contrast, fermions form tightly bound pairs in the BEC regime, and the literature~\cite{Leggett,PhysRevLett.91.030401} shows that the gap function now plays the role of the condensate wavefunction of the composite bosons and is proportional to the square root of the composite-boson density. Therefore, the gap function and density are tied to each other in the BEC regime and vanish together in the vortex core, eliminating those features from a normal Fermi gas in the cores of MQVs. Therefore, MQVs in Fermi superfluids in the BCS-BEC crossover reveal rich physics from pairing of fermions.

We emphasize that the spherical bubble trap has the following advantages for realizing and probing MQVs in cold-atom systems. First, the compact geometry of a sphere provides a tight confinement of the Fermi superfluid to prevent the atoms from escaping to spatial infinity due to high angular velocity and avoids unnecessary distortion of the density profile from the introduction of additional confining potentials. As explained below Eq.~\eqref{eq:TwL}, high angular velocity helps stabilize the MQVs, so tight confinement of the atoms is important. Second, the azimuthal symmetry of a sphere rotating about a fixed axis pins the pair of vortices at the two poles and disfavor their decay into configurations which break the symmetry. We caution that imperfections or fluctuations in experiments may violate the azimuthal symmetry and break the MQVs into clusters of vortices, as discussed in Refs.~\cite{Tomishiyo23,White23} for bosonic superfluids. To account for possible violation of the azimuthal symmetry in the calculations, sectors of the BdG equations with different values of $m$ will couple to each other to account for the azimuthal variation. The generalization will complicate the numerical evaluation and demand more resources for its investigation. Third, as discussed in Ref.~\cite{He22}, the BCS-BEC crossover of a Fermi superfluid on a thin spherical shell can be induced by tuning the size of the sphere or the interactions, allowing more controls in experiments to explore the physics of MQVs.

\section{Conclusion}\label{sec:conclusion}
The vortex solutions from the BdG equation reveal the structures of MQVs of a Fermi superfluid in a spherical bubble trap approximated by a thin shell  across the BCS-BEC crossover. The agreement of the single-vorticity vortex structure with that in the planar geometry reflects the local nature of the vortex. Nevertheless, in the cores of higher-vorticity vortices in the BCS regime, a density peak emerges due to an accumulation of a normal Fermi gas as the gap vanishes, and reversed circulation of the current occurs due to the in-gap states carrying finite angular momentum. Moreover, the number of in-gap state branches is topologically related to the vorticity of the vortex. Our results of MQVs in Fermi superfluids in spherical bubble traps thus demonstrate interesting interplay between geometry, many-body physics, and topology.

\begin{acknowledgments} 
Y. H. was supported by the NNSF of China (No. 11874272) and Science Specialty Program of Sichuan University (No. 2020SCUNL210). C. C. C. was partly supported by the NSF Grand No. PHY-2310656 and thank the hospitality of the KITP supported by NSF Grant No. PHY-1748958 and and PHY-2309135.
\end{acknowledgments}

%

\end{document}